\documentclass[aps,prl,twocolumn,showpacs]{revtex4}

\usepackage{graphicx}
\usepackage{amsmath,amssymb}
\usepackage{natbib}
\usepackage{hyperref}

\begin{document}
\preprint{}

\title{Entropic Stabilization of Tunable Planar Modulated
  Superstructures}
\author{Michael Engel}
\email{engelmm@umich.edu}
\affiliation{Department of Chemical Engineering, University of
  Michigan, Ann Arbor, Michigan 48109-2136, USA}
\date{\today}

\begin{abstract}
Self-assembling novel ordered structures with nanoparticles has
recently received much attention. Here we use computer simulations to
study a two-dimensional model system characterized by a simple
isotropic interaction that could be realized with building blocks on
the nanoscale. We find that the particles arrange themselves into
hexagonal superstructures of twin boundaries whose superlattice vector
can be tuned reversibly by changing the temperature. Thermodynamic
stability is confirmed by calculating the free energy with a
combination of thermodynamic integration and the Frenkel-Ladd
method. Different contributions to the free energy difference are
discussed.
\end{abstract}


\maketitle

The synthesis of nanoparticles and colloids with various types of
interactions~\cite{Bishop:2009yg} and shapes~\cite{Glotzer:2007eu} has
advanced rapidly. This opens up the possibility to use them as
building blocks for self-assembling novel materials `from bottom
up'. For future applications, ordered structures are especially
interesting because they have unique photonic and electronic
properties. Furthermore, it is desirable to have a large pool of
different types of crystal structures to choose from. Structure
formation of nanoparticles is guided by the same principles as
crystallization of atoms despite the different character of the
interparticle forces. Like for the majority of chemical elements, the
most common ground states of nanoscopic building blocks are
close-packed lattices which maximize density and the number of
neighbor bonds. Different and more complex
mesocrystals~\cite{Song:2010gd} have been observed experimentally for
example in binary mixtures of colloids~\cite{Shevchenko:2006fp,
  Leunissen:2005jt}, cylindrical liquid crystalline
phases~\cite{Tschierske:2007sf}, and in simulations of hard
tetrahedra~\cite{Haji-Akbari:2009qr}. In general, however, the
relationship between particle shape and interaction on the one hand,
and the stabilized crystal structure on the other hand is not well
understood.

Among the most complex ordered phases are commensurately and
incommensurately modulated crystals~\cite{Bak:1982la}, which have a
basic structure and a superstructure in form of a spatial
modulation. Although frequently found on the atomic scale in the
bulk~\cite{vanSmaalen:2007aa,Janssen:2007aa} and as ordered structures
of noble gases in adsorbed layers~\cite{Abraham:1984uq,
  Persson:1992fk}, modulated phases have not been known to
self-assemble with nanoparticles. So far only one-dimensional
modulations have been forced upon a two-dimensional system of
colloids~\cite{Baumgartl:2004lr} and binary
hard-disks~\cite{Franzrahe:2007ao} using an external periodic
potential. The purpose of this work is two-fold: First, we show that
planar superstructures with a tunable modulation period form in a
simple model system without external potential. Second, we confirm the
thermodynamic stability of the superstructures with free energy
calculations that are sensible enough to distinguish different
modulations and analyze the origin of their entropic stabilization.

\begin{figure}
  \centering
  \includegraphics[width=0.90\columnwidth]{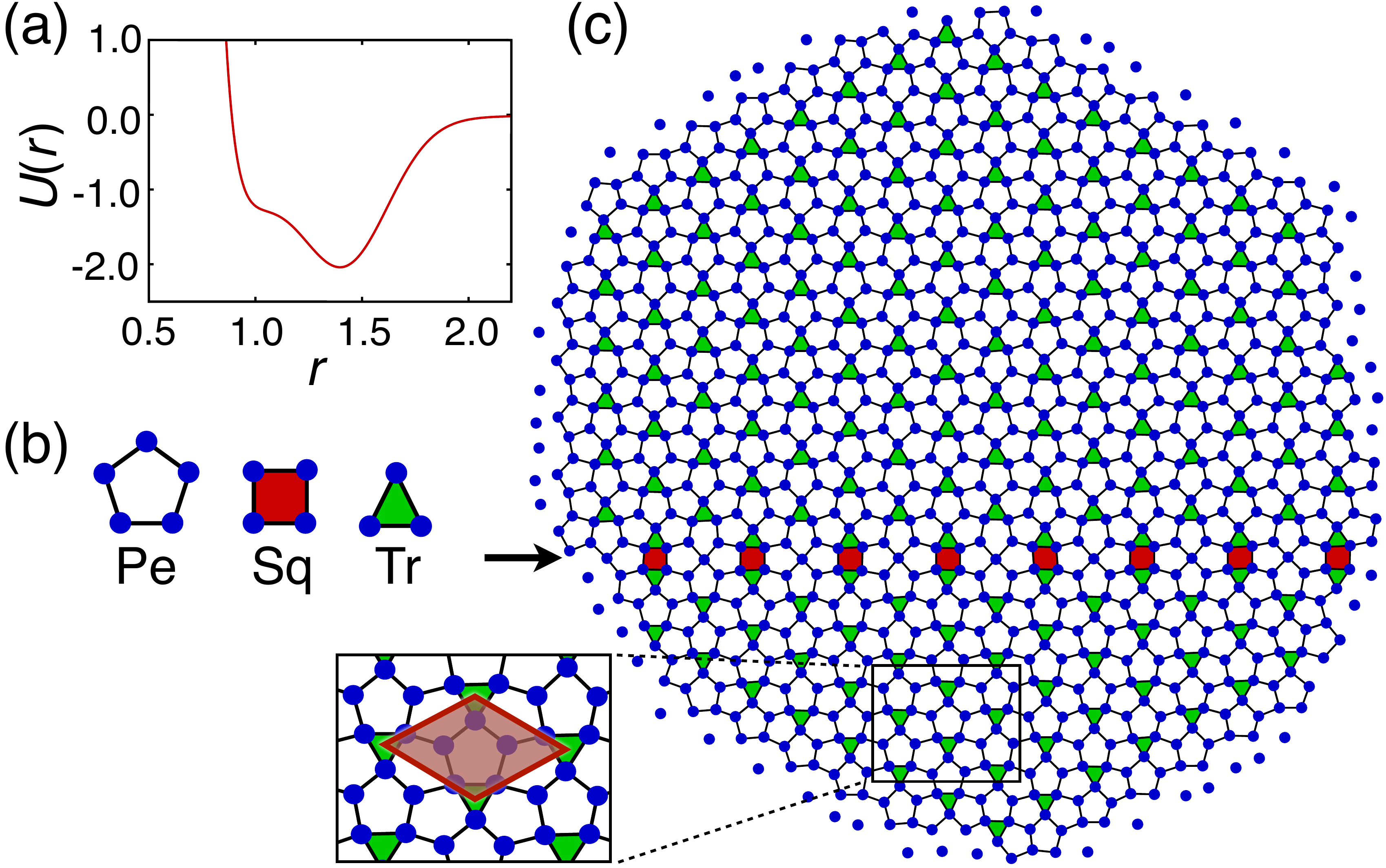}
  \caption{(color online) The interaction potential $U(T)$ with a
    broad minimum~(a) arranges the particles (disks) into (Pe)ntagons,
    (Sq)uares and (Tr)iangles~(b). (c)~The energetic ground state is
    Pe$_3$Tr. The inset outlines a unit cell. A twin boundary is
    marked with an arrow.\label{fig1}}
\end{figure}

Our model consists of identical particles interacting in two
dimensions via a Lennard-Jones-Gauss potential~\cite{Engel:2007hl}
with parameters $\sigma^{2}=0.042$, $\epsilon=1.8$, $r_{0}=1.42$. This
potential is isotropic and consists of a single minimum with a
shoulder (Fig.~\ref{fig1}(a)). Possible experimental realizations are
spherical macromolecules and micelles.  In the latter, a shoulder can
be caused by a repulsive hard core of carbon rings and an attractive
soft alkyl corona~\cite{Ziherl:2001fk,Zeng:2004uq}. We integrate the
particle motions with molecular dynamics (MD) computer simulations in
the $NPT$ ensemble at external pressure $P=0$. A potential cut-off at
$r=2.5$ and a thermostat for temperature control are used. Typical
simulation times are $10^6-10^9$ MD steps. Units are dimensionless and
fixed by the choice of the potential. An exception is the temperature
$T$, which is measured relative to a critical temperature
$T_C=0.35$. It has been shown in a previous study that the system
transforms into a square crystal at $T_C$ before it melts above
$T_M=0.40$~\cite{Engel:2010aa}.

A system of 1024 particles is slowly cooled down using open boundary
conditions to find the energetic ground state. Below $70\%$~$T_C$, the
system completely orders into a crystal with symmetry group $p3m1$
(Fig.~\ref{fig1}(c)). Since the particles are arranged in one triangle
tile (Tr) and three slightly deformed pentagons tiles (Pe) per unit
cell, we call this phase Pe$_3$Tr. Additional squares (Sq) tiles (with
two triangles attached) are present as defects. They arrange along a
line forming a twin boundary (TB) marked by an arrow in the
figure. The effect of the TB on the crystal structure is a flip of the
triangle tile orientations. We can measure the defect energy of a TB
by comparing the potential energy of a perfect Pe$_3$Tr crystal with
one that has a single TB. The energy change per unit length caused by
the presence of the TB is small compared to the average kinetic energy
at the critical temperature, but positive: $E/l=+0.13(1)$. This
confirms that the creation of the TB requires energy.

\begin{figure}
  \centering
  \includegraphics[width=1.0\columnwidth]{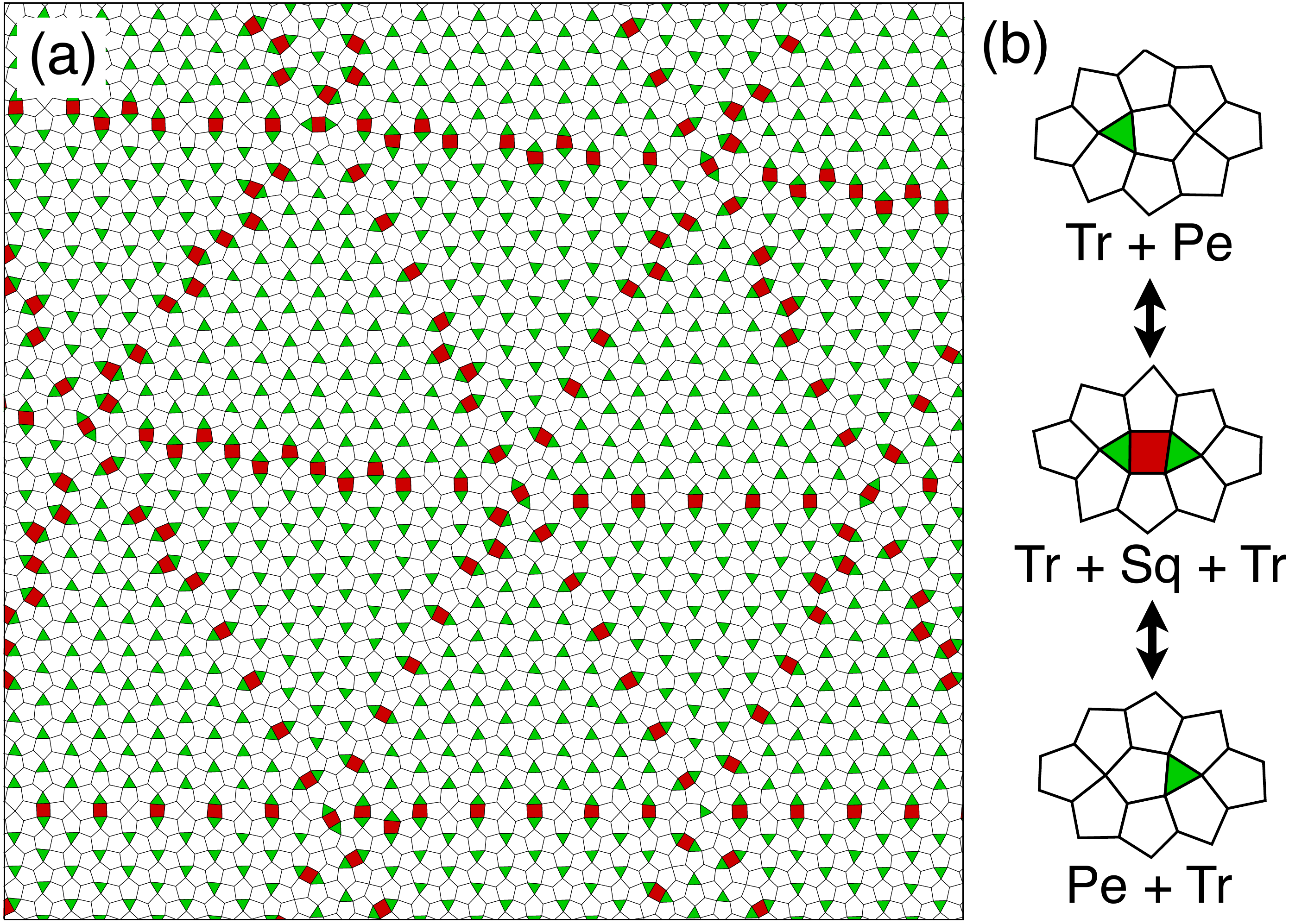}
  \caption{(color online) (a)~Ordering of twin boundaries into a
    hexagonal superstructure. Only the tiling, not the particles are
    shown. (b)~A structural rearrangement can transform a pentagon
    into a square plus triangle and induce a motion of twin
    boundaries. This collective motion is a phason mode.\label{fig2}}
\end{figure}

Next, we switch to a larger system with 10000 particles and study the
temperature dependence of the equilibrium crystal structure in more
detail. Fig.~\ref{fig2}(a) shows the system held for a long time at
$75\%$~$T_C$. We used a rapid quench to $T=0$ to remove thermal
noise. At the temperature $75\%$~$T_C$, TBs are found to appear in
equilibrium and arrange into a highly ordered hexagonal superstructure
(symmetry group $p6m$): (i)~The orientation of TBs is restricted to
three directions. (ii)~TBs are in average equally spaced. (iii)~TBs
form triple intersections only, double intersections never occur.
These observations are sufficient to identify the superstructure as an
interface modulated~\cite{Cummins:1990vl} or, alternatively, domain
wall incommensurate phase~\cite{Persson:1992fk}.

Given the simple pair interaction, the superstructure in
Fig.~\ref{fig2}(a) is surprisingly complex. The average periodicity
equals 33 nearest neighbor distances and the superstructure unit cell
contains approximately 700 particles. Furthermore, as will be shown
below, this superlattice constant can be tuned continuously by varying
the temperature. For this to happen there needs to be a way for the
TBs to propagate efficiently through the structure in form of a phason
mode~\cite{Bak:1982la}. The basic mechanism for such a motion is
depicted in Fig.~\ref{fig2}(b). Squares can change position in three
steps: (i)~The combination of a square and a neighboring triangle
transforms into a pentagon by moving the two particles of the common
edge outwards. As a result, an isolated triangle remains, which shares
edges with three pentagons. (ii)~This transformation is energetically
disfavorable and generates internal stress, which can be relaxed by
transforming another pentagon nearby into a combination of a square
and a triangle. (iii)~Subsequent propagation of all squares in a TB
induces a motion of this TB.

\begin{figure}
  \centering \includegraphics[width=1.0\columnwidth]{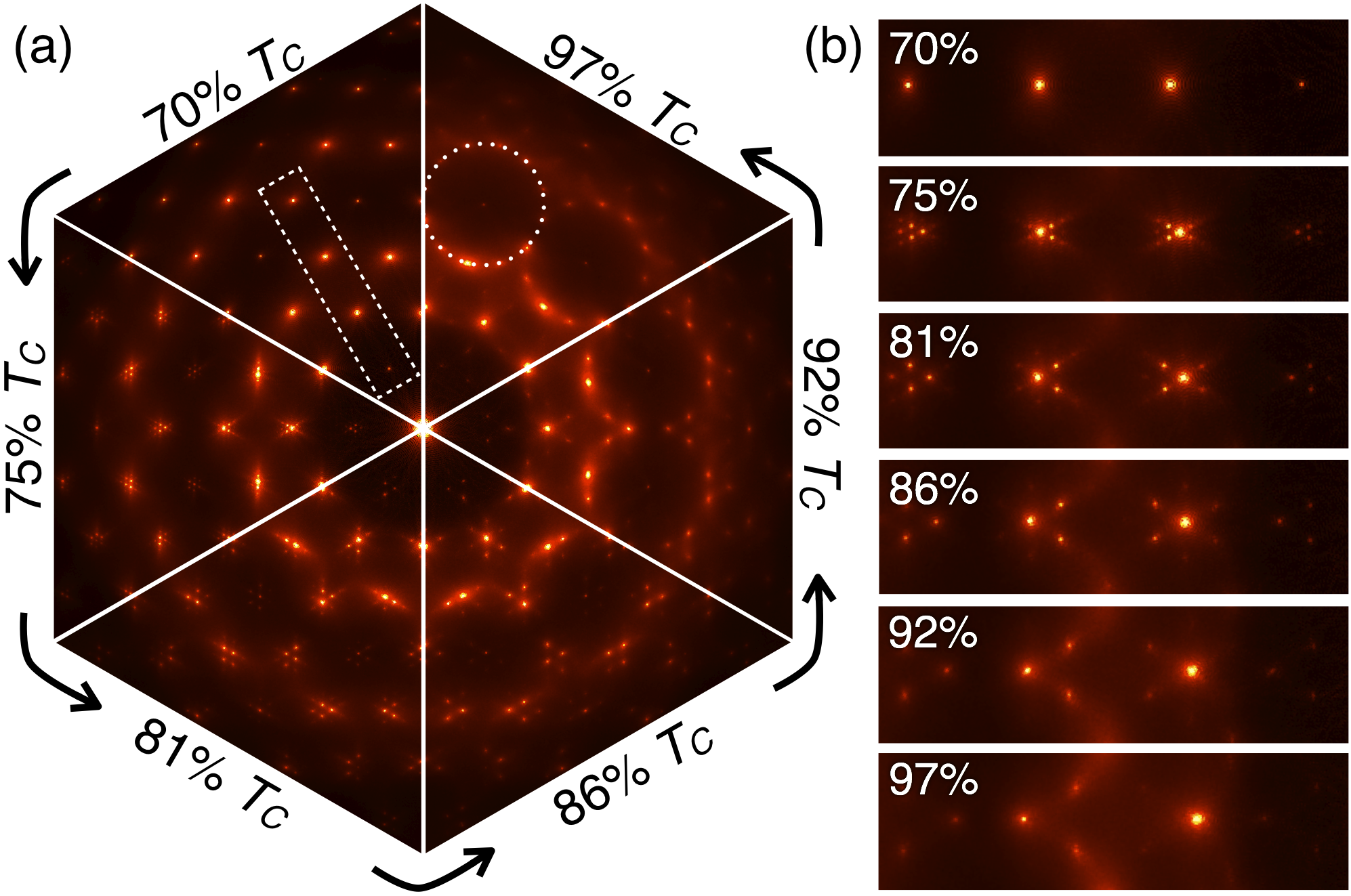}
  \caption{(color online) (a)~Each sector shows a diffraction image
    obtained by fast Fourier transform of the time averaged particle
    density. The temperature increases
    counter-clockwise. (b)~Magnified views of slices of the
    diffraction images (indicated by a dashed rectangle in (a)) show
    the increase of the superlattice wave vector.\label{fig3}}
\end{figure}

The hexagonal superstructure is best analyzed in Fourier
space. Diffraction images (DIs) at six different temperatures are
shown in Fig.~\ref{fig3}. Up to $70\%$~$T_C$ the Pe$_3$Tr crystal is
stable. The DI at this temperature consists of equally spaced, sharp
peaks. Only weak diffuse scattering caused by thermal motion is
present in the background. Above $70\%$~$T_C$, the peaks split up into
a strong central peak and weak satellite peaks as expected for a
modulated phase. The wave vector separating the satellite peaks
continuously increases with increasing temperature as a result of a
decreasing superlattice constant. A change in the rotational symmetry
of the DI is visible in Fig.~\ref{fig3}(a): The six-fold symmetry
present at low and intermediate temperatures gradually transforms into
a twelve-fold symmetry (generalized symmetry group $p12m$). Note that
at $97\%$~$T_C$, twelve peaks form two dodecagonal rings in the DI as
outlined by a dotted circle. This suggests that a dodecagonal
quasicrystal is stabilized close to the critical temperature. The
structure and dynamics of this quasicrystal has been studied
recently~\cite{Engel:2010aa}. The sequence of symmetry changes is
summarized as:
\begin{equation}\label{eq1}
p3m1\xrightarrow{0.7T_C}p6m\xrightarrow{0.97T_C}p12m
\xrightarrow{T_C}p4m\xrightarrow{T_M}E(2).
\end{equation}
While the first two phase transitions are second order, the transition
to the square crystal at $T_C$ and the melting transition at
$T_M=1.15T_C$ are first order~\cite{Engel:2010aa}. Snapshots of the
particle configurations at various temperatures and full diffraction
patterns can be found in the supplementary information.

\begin{figure}
  \centering
  \includegraphics[width=0.85\columnwidth]{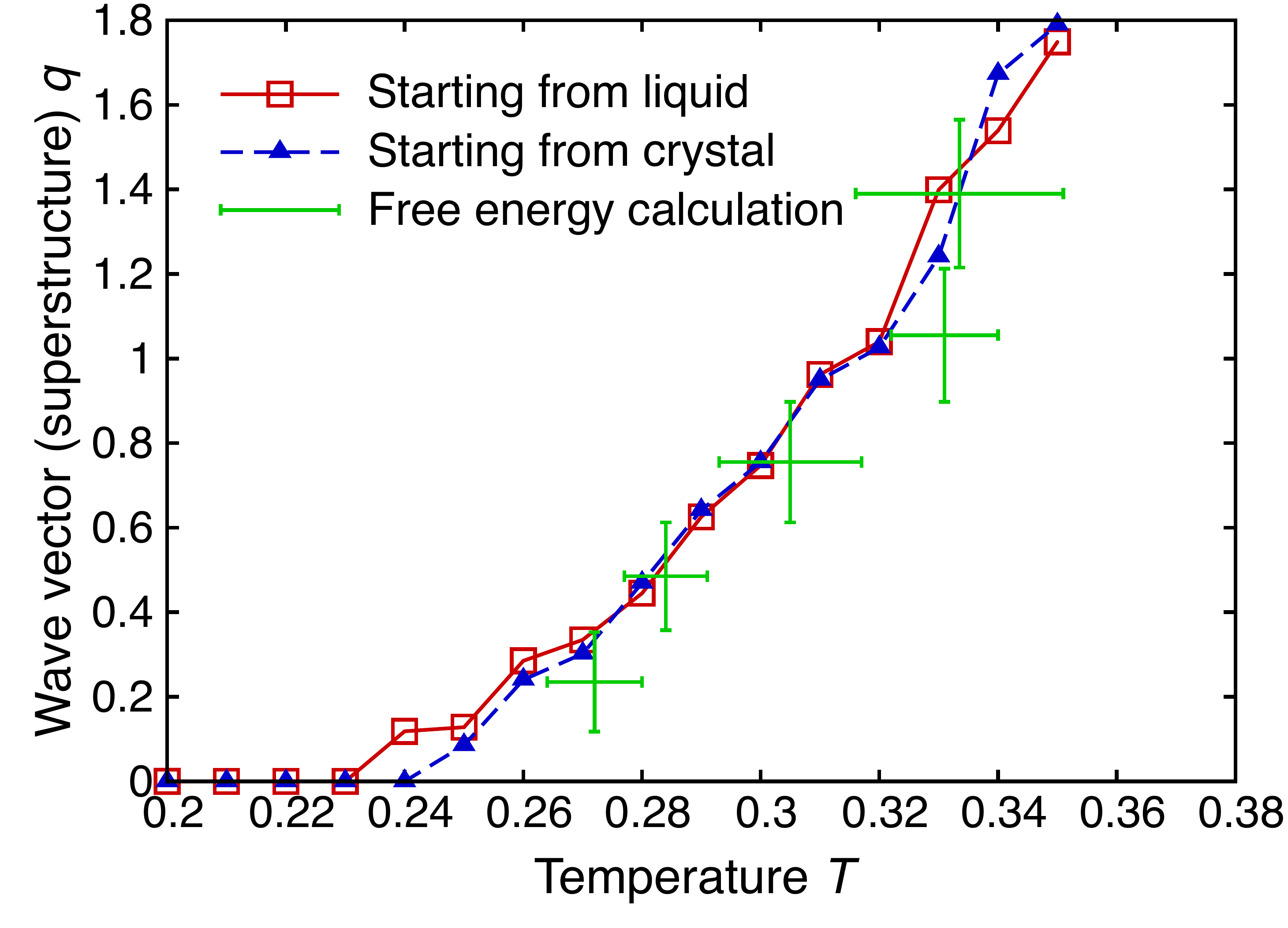}
  \caption{(color online) The wave vector of the superstructure grows
    linearly with temperature. No hysteresis is observed in molecular
    dynamics simulations. The growth of the superlattice wave vector
    is confirmed by free energy calculations.\label{fig4}}
\end{figure}

We determine the superlattice wave vector by measuring the distance
$q$ of two neighboring superlattice peaks. Simulations are started
from Pe$_3$Tr (stable at low $T$) and the fluid (stable at high $T$)
to test for hysteresis in the temperature behavior. The dependence of
the superlattice wave vector on temperature is plotted in
Fig.~\ref{fig4}. No hysteresis is present and the final configuration
is found to be independent of the starting condition, which is an
indication that thermodynamic equilibrium has been reached. The wave
vector of the superstructure increases roughly linearly between
$T=0.24\approx 70\%$~$T_C$ and $T_C$. Such a continuous increase of
the wave vector is typical for two-dimensional systems and is
characteristic for a `floating' modulation~\cite{Bak:1982la}.

The stabilization of the superstructures is analyzed in more detail by
calculating free energies. Two complementary techniques are used: the
Frenkel-Ladd method (FL) and thermodynamic integration (TI). The
absolute free energy can be obtained at all temperatures combining the
methods into a new method (FLTI): (i)~At low temperature
($<10\%$~$T_C$), the absolute free energy is determined with FL by
interpolating between the target structure and an Einstein crystal
with known free energy~\cite{Frenkel:1984zp}. (ii)~The free energy is
then extended to higher temperatures with TI by integrating the
potential energy $E(T)$ along a reversible path:
\begin{equation}\label{eq2}
F(T)=c(T_0)T-T\int_{T_0}^{T}\text{d}T'\frac{E(T')}{T'^2},
\end{equation}
where $T_0$ is a reference temperature and $c(T_0)$ an integration
constant that is determined from a comparison with FL. Using both
methods is necessary because FL cannot be applied at high
temperatures. The reason is the motion of the TBs, which makes
equilibration during the interpolation to the Einstein crystal very
slow due to the presence of particle rearrangements.

Boundary conditions are important for equilibrating the
superstructures. Open boundaries ensure that TBs can be created and
annihilated. In contrast, by switching to periodic boundary
conditions, the superlattice constant $q$ remains fixed because the
total number of TBs in the system cannot change. Using periodic
boundaries allows to simulate superstructures at temperatures where
they would otherwise be metastable with regard to a change in $q$.

We analytically construct superstructures with various values of $q$
and calculate the free energy $F(T,q)$ using FLTI and periodic
boundaries. Stable superstructures are characterized by $\partial
F/\partial q|_T=0$. Results from free energy calculation are
included in Fig.~\ref{fig4}. They agree well with the MD simulations
and confirm that the temperature dependence of the superstructures
wave vector observed in MD simulations.

The stabilization of the superstructures at elevated temperatures can
only be explained by a higher entropy in comparison to the Pe$_3$Tr
crystal. We now analyze the contributions to the free energy. In
general, the free energy of a classical $d$-dimensional system can be
written as a series:
\begin{equation}\label{eq3}
F(T)=\underbrace{E(0)-\tfrac{d}{2}Nk_BT\ln(T)+c'T}_{F_{\text{H}}(T)}+
\underbrace{\alpha T^2+\beta T^3+\ldots}_{F_{\text{A}}(T)}
\end{equation}
with new constants $c'$, $\alpha$, and $\beta$. This series splits up
into a `harmonic' part $F_{\text{H}}(T)$ and an `anharmonic' part
$F_{\text{A}}(T)$. The harmonic part corresponds to the free energy of
the system in a linear approximation, where the dynamics is completely
specified by the dynamical matrix, or alternatively, by the zero
temperature phonon density of state. The constant $c'$, and therefore
$F_{\text{H}}$, can be determined from the FL results at low
temperature. The anharmonic part contains higher-order effects and has
a more complicated temperature dependence. We write $F_{\text{A}}$
with the help of Eq.~(\ref{eq2}) in a form that can directly be used
in simulation:
\begin{equation}\label{eq4}
F_{\text{A}}(T)=-T\int_{0}^{T}\text{d}T'\frac{E(T')-
  E(0)-\tfrac{d}{2}Nk_BT'}{T'^2}.
\end{equation}
Note that since the integral does not diverge for small temperatures,
we can now take $T=0$ as the lower limit.

\begin{figure}
  \centering
  \includegraphics[width=0.85\columnwidth]{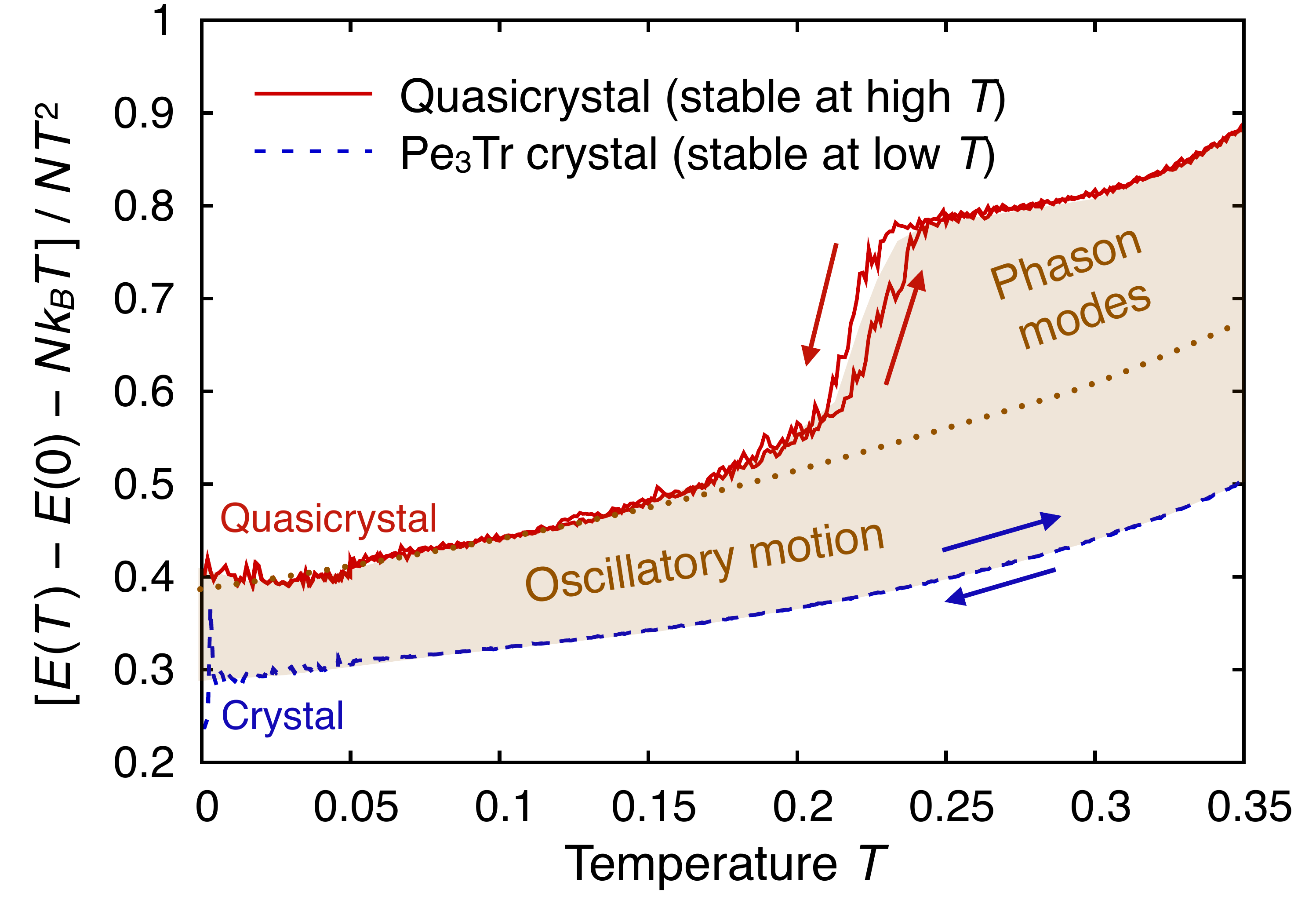}
  \caption{(color online) Analysis of the entropic stabilization of
    the quasicrystal and the superstructures. The anharmonic part of
    the free energy per particle, $F_A(T)/N$, can be obtained by
    integrating $[E(T)-E(0)-Nk_BT]/NT^2$, plotted in the figure. The
    dashed line splits contributions originating from oscillatory
    (phonon) motion and the phason modes.\label{fig5}}
\end{figure}

In order to study the temperature dependence of $F_{\text{A}}$, we
plot the integrand of this equation for the low temperature phase
Pe$_3$Tr (CR) and the quasicrystal (QC) stable at high temperature in
Fig.~\ref{fig5}.  The shaded area between the curves corresponds to a
free energy difference of the two phases.

The free energy difference at the critical temperature,
\begin{equation}
\Delta F=[F_{\text{QC}}(T_C)-F_{\text{QC}}(0)]-
       [F_{\text{CR}}(T_C)-F_{\text{CR}}(0)],
\end{equation}
has three contributions: (1)~Harmonic particle motion in the local
potential energy minima (HM). (2)~Anharmonic particle motion in the
local potential energy minima (AM). These two parts vary smoothly with
temperature. (3)~Particle rearrangements as part of the phason modes
(PH), which correspond to the TB motion, but can also be pentagon
rotations~\cite{Engel:2010aa}. While the first contribution is equal
to $T_C\Delta c'$, the last two are calculated from the shaded area in
Fig.~\ref{fig5}. Notice that phason modes only contribute after they
are activated at around $70\%$ $T_M$. We estimate their contribution
by extrapolating the low-temperature data for the quasicrystal as
indicated by the dashed curve. The results are: $\Delta
F_{\text{HM}}/N=0.042$, $\Delta F_{\text{AM}}/N=0.017$, $\Delta
F_{\text{PH}}/N=0.010$. The majority of the free energy difference
comes from oscillatory (phonon) motion. Interestingly, phason modes do
not contribute significantly.

In conclusion, we have introduced a model system that stabilizes a
continuum of modulated superstructures. The interaction potential is
simple enough that it could in principle be realized for example with
attractive nanoparticles. Complex two-dimensional structures are now
feasible experimentally: Only recently complex tunable two-dimensional
binary molecular networks have been fabricated~\cite{Huang:2010fk}. In
another work, it has been demonstrated that binary nanoparticles can
be assembled at the liquid-air interface into highly ordered
crystals~\cite{Dong:2010ud}. However, even without direct experimental
realization, the simulation techniques and theoretical ideas outlined
in this work are of interest. The combination of the Frenkel-Ladd
method and thermodynamic integration is a powerful tool that allows to
calculate free energies with very high precision ($<10^{-4}$). In our
case it was possible to compare superstructures with different wave
vectors and distinguish the free energy contributions from oscillatory
(phonon) particle motion and phason modes.

Support from the Deutsche Forschungsgemeinschaft (EN 905/1-1) is
gratefully acknowledged.

\bibliography{superstructure}

\includegraphics[width=\textwidth]{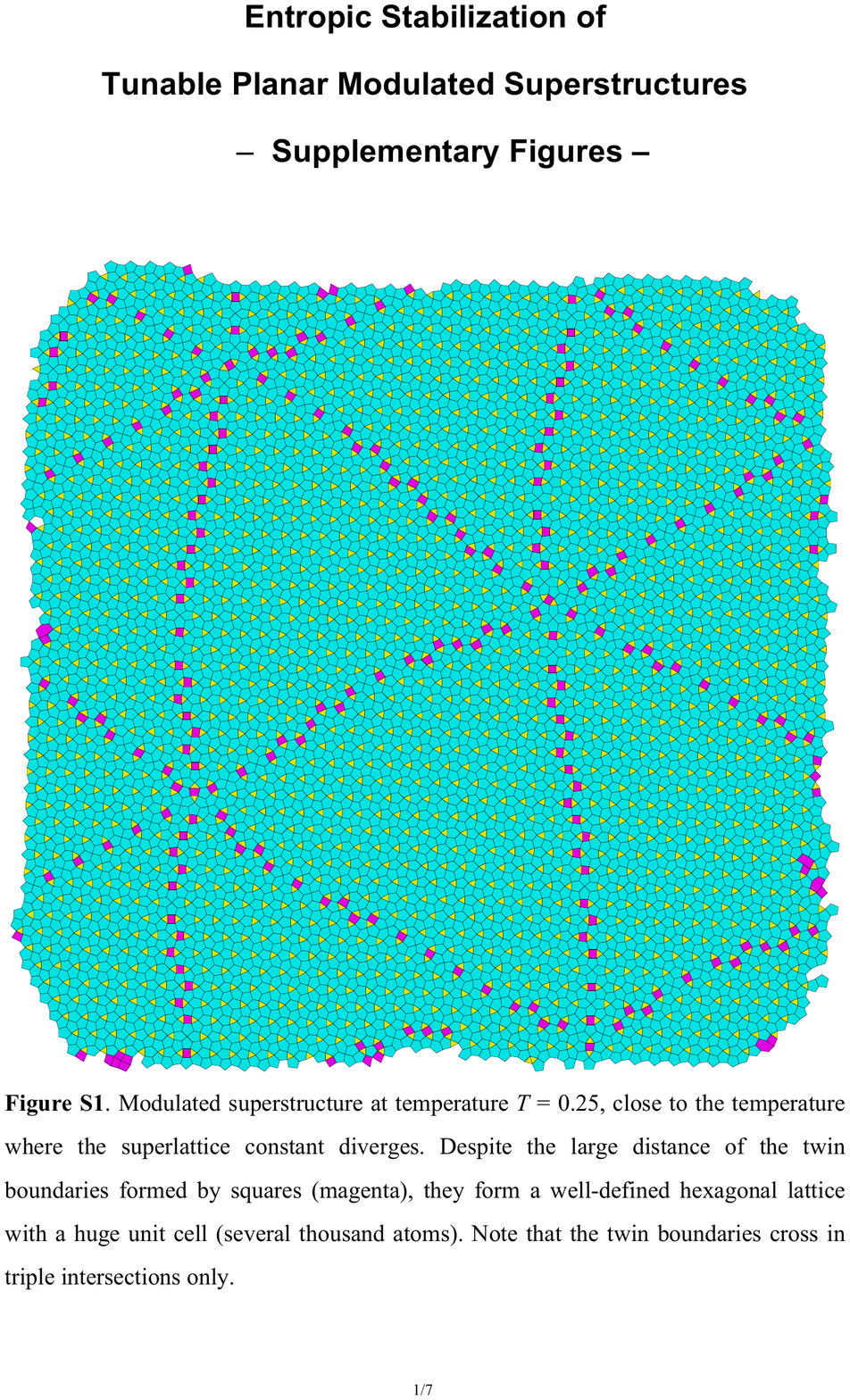}\\
\\

\includegraphics[width=\textwidth]{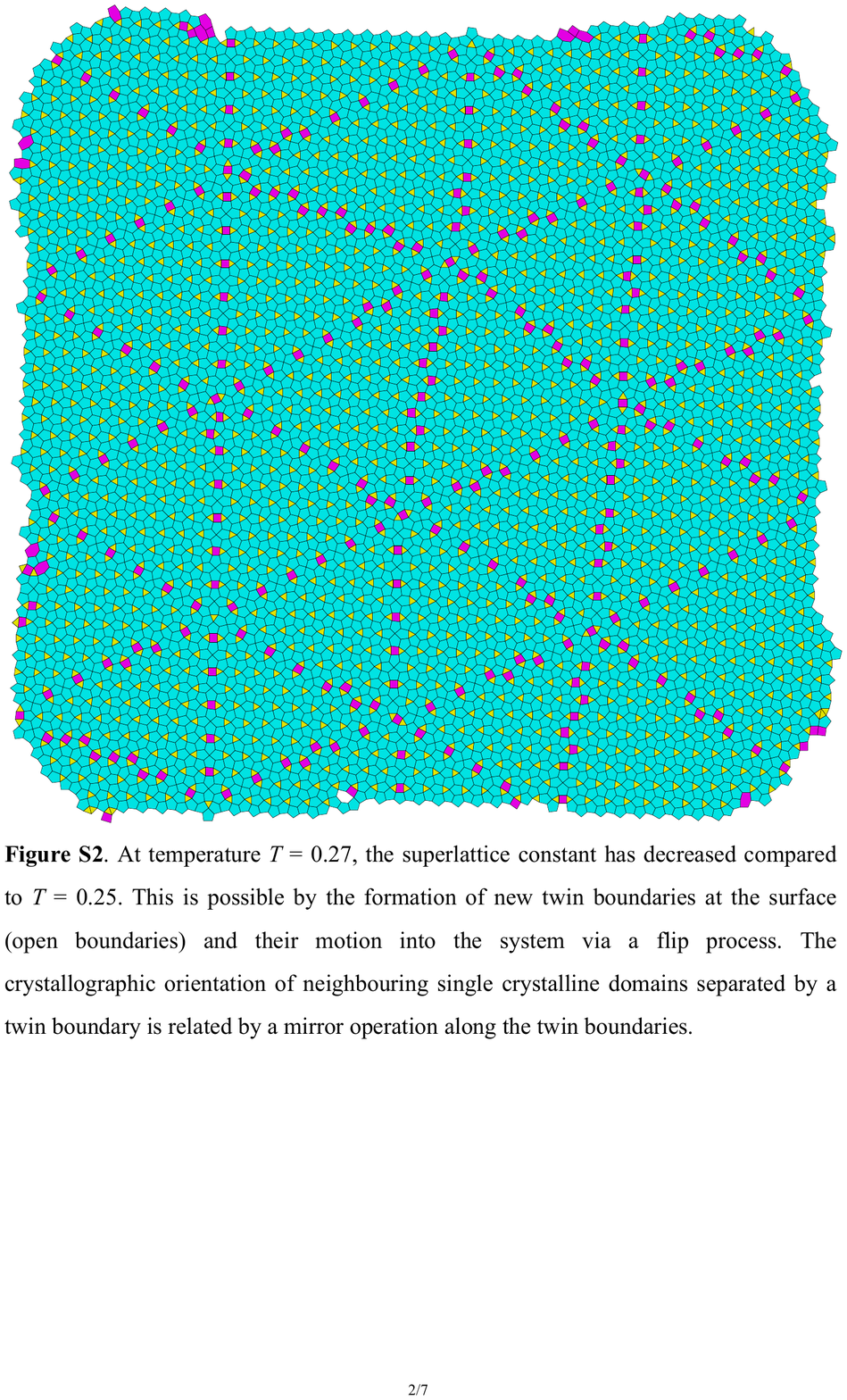}\\
\\

\includegraphics[width=\textwidth]{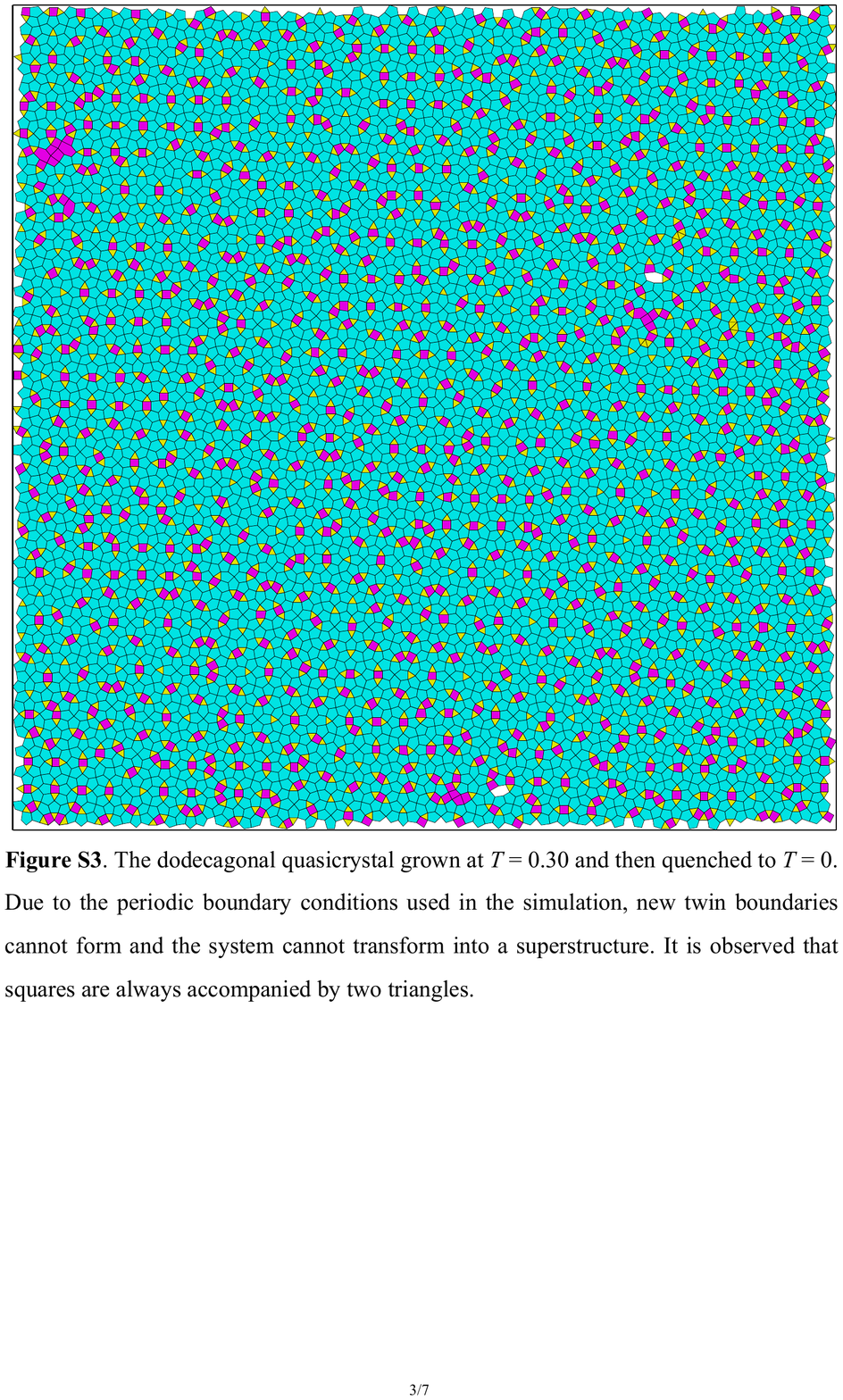}\\
\\

\includegraphics[width=\textwidth]{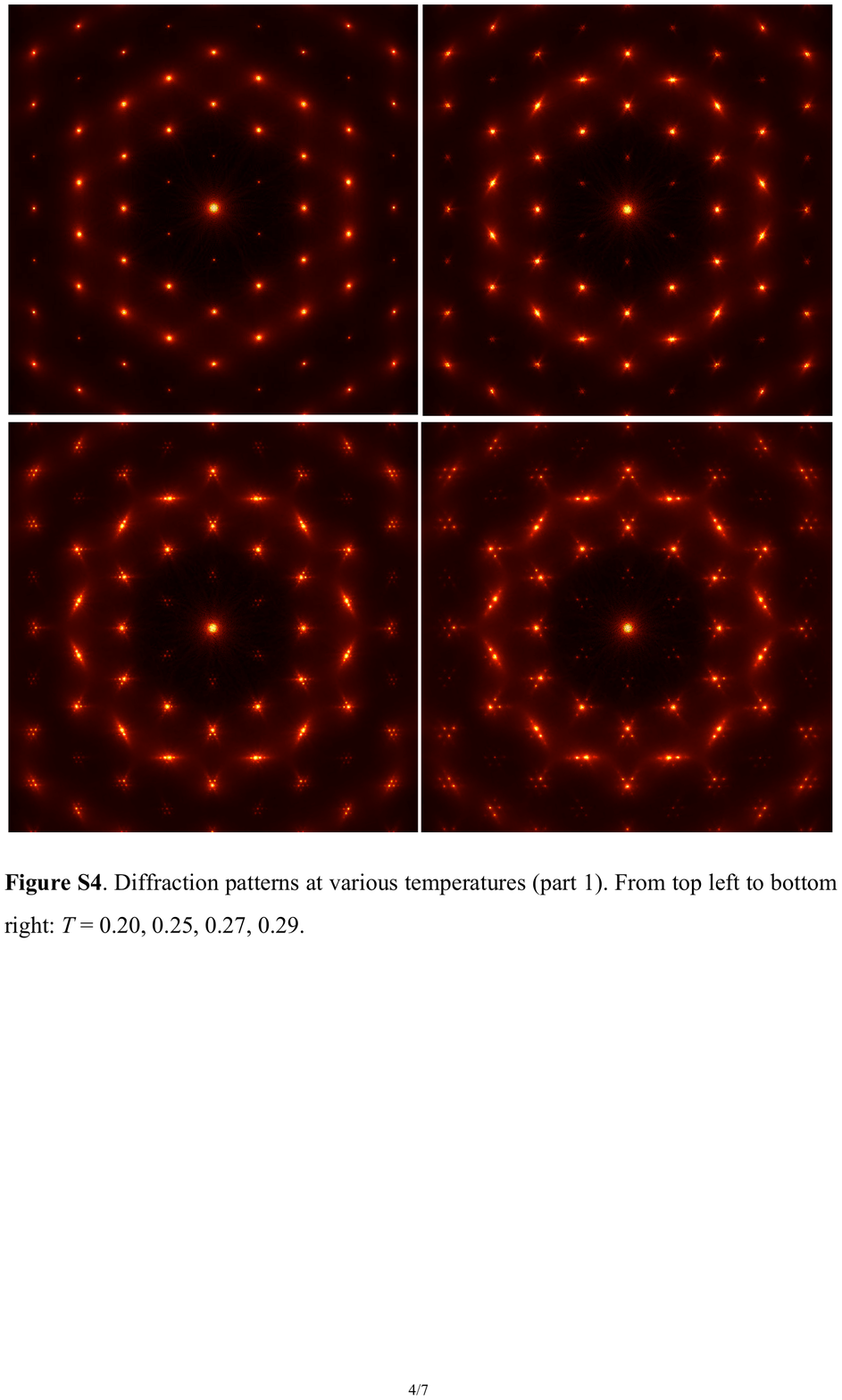}\\
\\

\includegraphics[width=\textwidth]{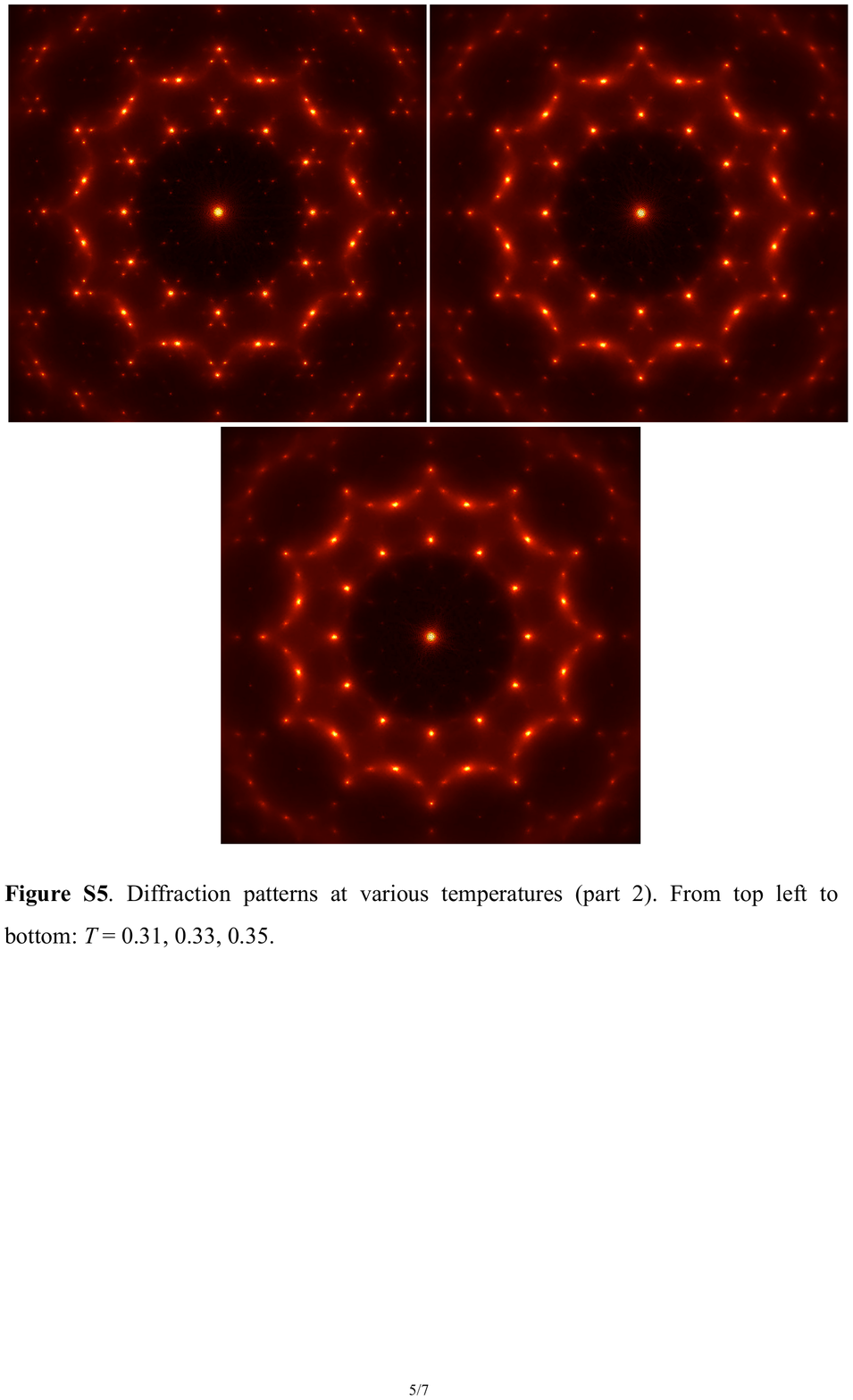}\\
\\

\includegraphics[width=\textwidth]{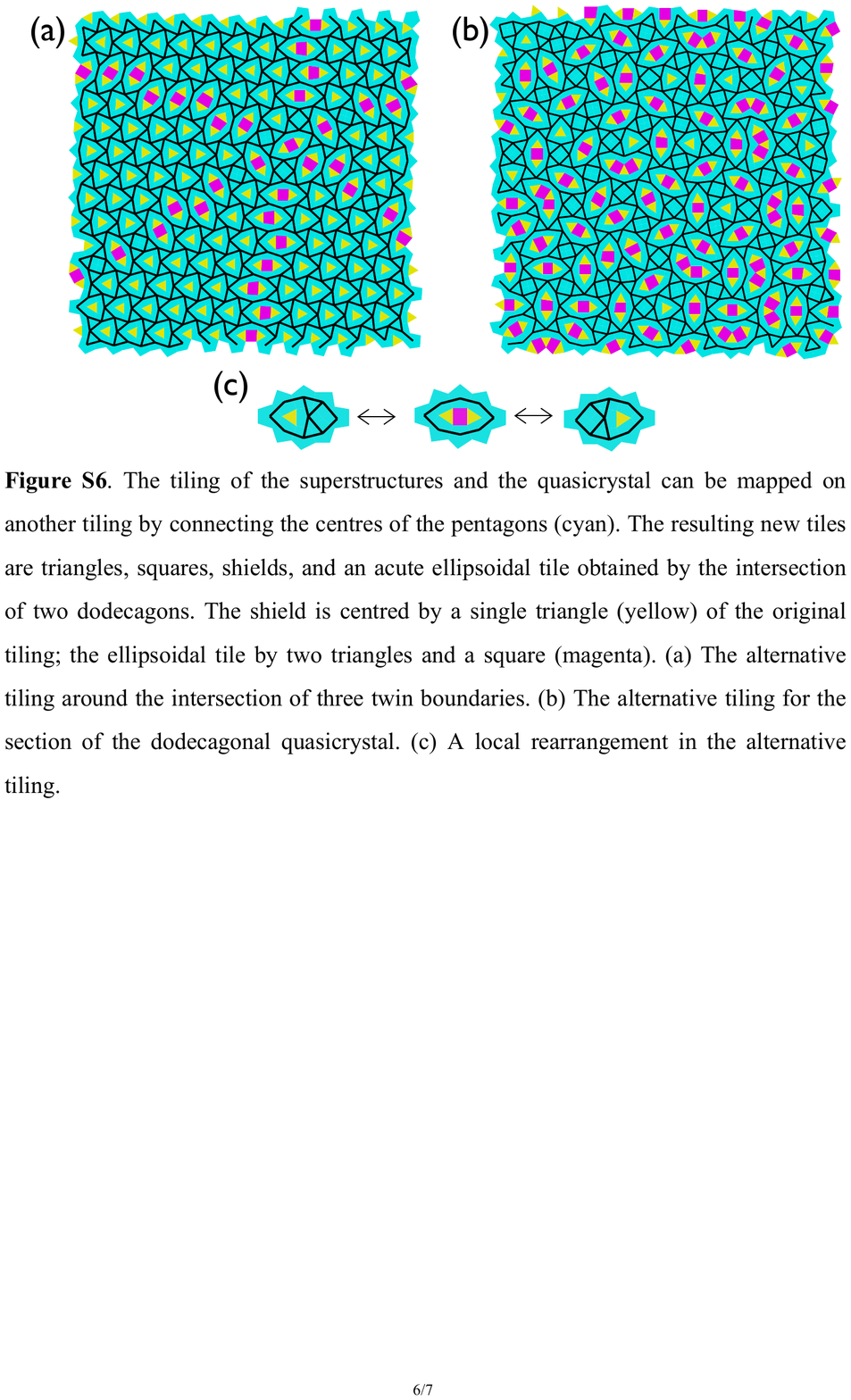}\\
\\

\includegraphics[width=\textwidth]{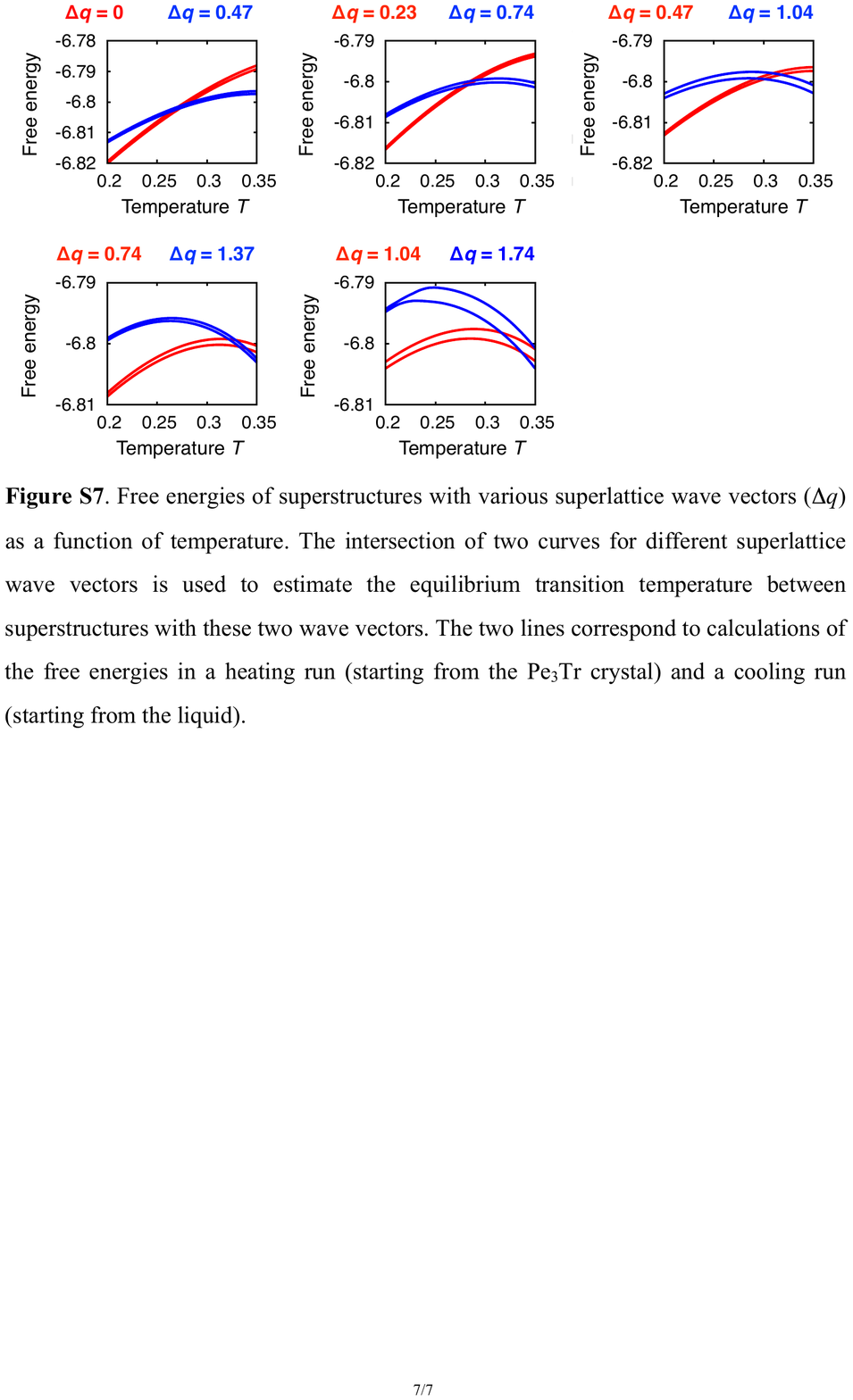}\\

\end{document}